# Variations in Abundance Enhancements in Impulsive Solar Energetic-Particle Events and Related CMEs and Flares

**Donald V. Reames[1], Edward W. Cliver[2], and Stephen W. Kahler[3]**

[1]Institute for Physical Science and Technology, University of Maryland, College Park, MD 20742-2431 USA, email: dvreames@umd.edu

[2]Space Vehicles Directorate, Air Force Research Laboratory, Sunspot, NM 88349, USA, email: ecliver@nso.edu

[3]Air Force Research Laboratory, Space Vehicles Directorate, 3550 Aberdeen Avenue, Kirtland AFB, NM 87117, USA, email: stephen.kahler@kirtland.af.mil

**Abstract** We study event-to-event variations in the abundance enhancements of the elements He through Pb for Fe-rich impulsive solar energetic-particle (SEP) events, and their relationship with properties of associated coronal mass ejections (CMEs) and solar flares. Using a least-squares procedure we fit the power-law enhancement of element abundances as a function of their mass-to-charge ratio $A/Q$ to determine both the power and the coronal temperature (which determines $Q$) in each of 111 impulsive SEP events identified previously. Individual SEP events with the steepest element enhancements, *e.g.* $\sim(A/Q)^6$, tend to be smaller, lower-fluence events with steeper energy spectra that are associated with B- and C-class X-ray flares, with cooler (~2.5 MK) coronal plasma, and with narrow (<100°), slower (<700 km s$^{-1}$) CMEs. On the other hand, higher-fluence SEP events have flatter energy spectra, less-dramatic heavy-element enhancements, *e.g.* $\sim(A/Q)^3$, and come from somewhat hotter coronal plasma (~3.2 MK) associated with C-, M- and even X-class X-ray flares and with wider CMEs. Enhancements in $^3$He/$^4$He are uncorrelated with those in heavy elements. However, events with $^3$He/$^4$He ≥ 0.1 are even more strongly associated with narrow, slow CMEs, with cooler coronal plasma, and with B- and C-class X-ray flares than are other Fe-rich impulsive SEP events with smaller enhancements of $^3$He.

# 1. Introduction

Historically, the relative abundances of the chemical elements have been one of our most powerful tools in identifying the source of a variety of energetic ion populations seen throughout the heliosphere and in discovering physical processes involved in their acceleration and transport (*e.g.* Reames 1999). Since the pioneering work of Meyer (1985) it has been known that the abundances of elements in large "gradual" solar energetic-particle (SEP) events are closely related to the corresponding abundances of elements in the solar corona. SEPs in gradual events are accelerated, in proportion to the ambient "seed population," by shock waves, driven out from the Sun by fast, wide coronal mass ejections (CMEs). These abundances in gradual SEP events contrast sharply with the spectacular abundance enhancements of the smaller, more numerous, "impulsive" SEP events that have 1000-fold enhancements of both $^3$He/$^4$He and heavy elements, *i.e.* (Z≥50)/O, produced during acceleration in solar flares and jets (for a recent review of gradual and impulsive SEP events see Reames, 2013). The distinction of gradual and impulsive SEP events involves a wide variety of evidence (*e.g.* Reames, 1990, 1995a 1995b, 1999, 2002, 2013; Kahler, 1992, 1994, 2001; Gosling, 1993; Reames, Meyer, and von Rosenvinge, 1994; Lee, 1997, 2005; Mason Mazur, and Dwyer, 1999; Tylka, 2001; Gopalswamy *et al.*, 2002; Ng, Reames, and Tylka, 2003; Desai *et al.*, 2003, 2004, 2006; Slocum *et al.*, 2003; Cliver, Kahler, and Reames, 2004; Tylka, *et al.*, 2005; Tylka and Lee, 2006; Cliver and Ling, 2007, 2009; Cohen *et al.*, 2007; Leske *et al.*, 2007; Ng and Reames, 2008; Sandroos and Vainio, 2009; Rouillard *et al.*, 2011, 2012; Wang *et al.*, 2012). The relatively recent expansion of measurements to elements throughout the rest of the periodic table above Fe (Reames, 2000; Mason *et al.*, 2004; Reames and Ng, 2004; Mason, 2007; Reames, Cliver, and Kahler, 2014) has contributed significantly to the discrimination and characterization of impulsive SEP events.

The distinctive enhancements in $^3$He/$^4$He, characteristic of impulsive SEP events, have been ascribed to resonant wave-particle interactions in the flare plasma, most recently to electromagnetic ion cyclotron waves produced near the gyrofrequency of $^3$He by copious electrons beams streaming out from the flare (Temerin and Roth, 1992; Roth and Temerin, 1997; see also Liu, Petrosian, and Mason 2006), but the smooth rise of element abundance enhancements with increasing *A/Q* is not fit well by a resonant model. The fact that heavy-element enhancements are uncorrelated with enhancements in $^3$He/$^4$He (*e.g.* Mason *et al.*, 1986; Reames, Meyer, and von Rosenvinge, 1994) suggests the operation of more than one physical mechanism. Recently, the heavy-element enhancements, rela-tive to coronal abundances, and their strong power-law dependence on *A/Q* of the ions in impulsive SEP events, have been linked theoretically to the physics of magnetic reconnection regions from which the ions escape (Drake *et al.*, 2009; Knizhnik, Swizdak, and Drake, 2011; Drake and Swizdak, 2012).

Regarding the solar associations of gradual and impulsive SEP events, Kahler *et al.* (1984) established a clear (96%) association of the *gradual* events with fast, wide CMEs. However, Kahler, Reames, and Sheeley (2001) showed that some classic *impulsive* SEP events were associated with narrow CMEs and Yashiro *et al.* (2004) found that at least 28-39% of impulsive SEPs had associations with CMEs as well as having their previously-known association with solar flares and type III radio bursts. These associations have tied impulsive SEP events to the mechanism described by the theory of jets (Shimojo and Shibata, 2000) where emerging magnetic flux reconnects on open field lines allowing easy escape of SEPs and of plasma, the CME (Shimojo and Shibata, 2000; Kahler, Reames, and Sheeley, 2001; Reames, 2002; Nitta *et al.*, 2006; Wang, Pick, and Mason, 2006; Moore *et el.*, 2010; Archontis and Hood, 2013).

In a recent paper, Reames, Cliver, and Kahler (2014, hereinafter Paper 1) found that a large fraction (69%, perhaps more) of the 111 impulsive SEP events they studied was associated with CMEs; 68% were associated with Hα flares. Furthermore, element abundances in the events seemed to vary with broad properties of the associated CMEs, such as their speed and angular width. However, these authors mainly studied the average over all events of the strong power-law enhancement of abundances as a function of *A/Q* of the elements. The average enhancement was found to vary as the 3.6 power of *A/Q*, where the value of *Q* was determined at a coronal temperature near 3 MK. This temperature region, where *A/Q* for Ne exceeds that for Mg and Si in inverted order of *Z*, rather uniquely explains the unusually large enhancement in Ne/O which exceeds that of Mg/O and Si/O. However, these results beg additional questions. How does this power of *A/Q* vary from one SEP event to another? Is the event temperature defined by SEP ionization always ~3 MK? Do these parameters in different events vary with the X-ray intensity or with the speed or angular width of the associated CME? Do $^3$He-rich SEP events follow the pattern of other Fe-rich events?

The present paper may be considered as a sequel to Paper 1 where we now examine *each* of the 111 individual events listed therein to determine the most probable power of the enhancement *vs. A/Q* and the best value of source plasma temperature that determines *Q*. Given these properties we study their event-to-event variations and their possible dependence upon properties of the associated CMEs and flares or their link to other properties of the SEPs themselves, such as $^3$He/$^4$He. He, with no isotope specified, means $^4$He throughout this paper.

The SEP observations in this paper were measured by the *Low Energy Matrix Telescope* (LEMT: von Rosenvinge *et al.*, 1995) on the *Wind* spacecraft which measures elements from He through about Pb in the energy region from about 2 – 20 MeV amu$^{-1}$, identifying and binning the major elements from He to Fe onboard at a rate up to about $10^4$ particles s$^{-1}$. Instrument resolution and aspects of the processing have been shown and described elsewhere (Reames *et al.*, 1997; Reames, Ng, and Berdichevsky, 2001; Reames, 2000; Reames and Ng, 2004). Element resolution in LEMT from C–Fe, as a function of energy, has been shown recently and abundances in gradual events have been analyzed by Reames (2014). Typical resolution of He isotopes in LEMT was shown by Reames *et al.* (1997) and resolution of elements with *Z* >26 by Reames (2000). LEMT response was calibrated with accelerator beams of O, Fe, Ag, and Au before launch (von Rosenvinge *et al.*, 1995).

CME observations listed in Paper 1 were derived from the *Large Angle and Spectrometric Coronagraph* (LASCO; Brueckner *et al.*, 1995) on board the *Solar and Heliospheric Observatory* (SOHO). We also use peak intensities of 1–8 Å X-ray from *the Geostationary Operational Environmental Satellite* (GOES) X-ray Spectrometer (see http://spidr.ngdc.noaa.gov/spidr/index.jsp).

Our basic procedure for associating solar and SEP events was described in paper I. We first searched the Wind/Waves low-frequency (20 kHz – 13.8 MHz) radio data (Bougeret *et al*., 1995; http://www-lep.gsfc.nasa.gov/waves/) for candidate type III bursts for times from 1 h to 8 h before SEP event onset. The association of impulsive SEP events and type III radio bursts has been known for many years (*e.g.* Reames and Stone 1986). We found reasonable candidate type IIIs for 95 of the 111 impulsive SEP events. In identifying these type IIIs, we gave preference to those associated with Wind/3DP 1 to 300 keV solar electron events for the years 1995-2005 (Wang et al. 2012). Of the 81 type III candidates we identified from 1995-2005, 59 (73%) had an associated electron event. The type IIIs led the electron event onsets by a median value of 5 minutes. These type III onset times are tabulated in paper I and the procedure for identifying associated CMEs are described therein. The timing of the type III burst was used to identify the GOES 1-8 Å soft X-ray burst for each event. The median delay of the onset time of the type III emission relative to the onset of the associated soft X-ray burst was 3 minutes (range of -15 to +5 min).

D. V. Reames, E. W. Cliver, S. W. Kahler

## 2. Fitting the Power-Law Enhancements

### 2.1 Ionization and *A/Q vs.* Temperature

In order to plot the enhancement *vs. A/*Q, we need to determine the ionization state, *Q,* of each of the elements, which depends upon the plasma temperature in each event. Here we follow the logic of Reames, Meyer, and von Rosenvinge (1994) and Paper 1. Overall, there is a pattern of abundances in impulsive SEP events where He, C, N, and O are unenhanced in most events while Ne, Mg, and Si have similar enhancements, and Fe and heavier elements have increasingly greater enhancements. As in Paper 1, we have plotted *A*/Q as a function of equilibrium coronal temperature in Figure 1.

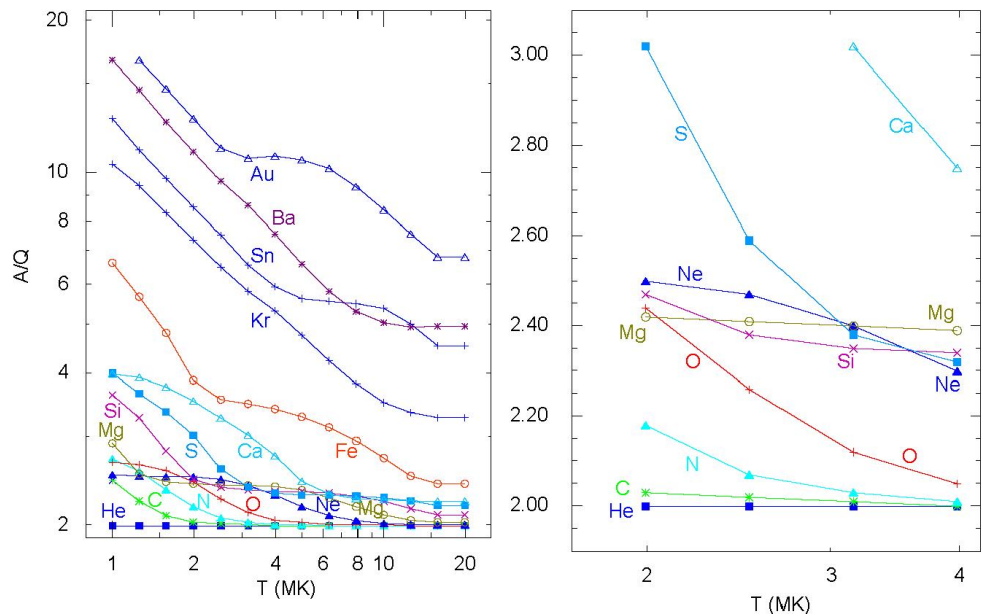

**Figure 1.** *A/Q* is shown as a function of equilibrium temperature for several elements from 1–20 MK (left panel) and in an enlarged region for low *Z* from 2–4 MK (right panel). Elements below Fe are from Arnaud and Rothenflug (1985), Fe from Arnaud and Raymond (1993) and sample elements in the high-*Z* region from Post *et al.* (1977). The figure is adapted from Paper 1.

In the temperature range near 3 MK, the elements He, C, N, and O are fully ionized, or nearly so, the elements Ne, Mg, and Si pass through He-like atomic states with only 2 orbital electrons and $Q_{Fe} \approx 15$. The region just below 3 MK is especially interesting since *A/*Q is higher for Ne than for Mg and Si. This inversion of *Z* ordering is seen in the pattern of enhancements. Note that below ~2 MK, *A/Q* for O exceeds that of Ne, and above ~6 MK, O and Ne both have *A/Q*=2, in both cases it would be hard to understand the observed enhancements in Ne/O.

Typical variation with temperature of the enhancement *vs. A/Q* is shown for two events in Figure 2. Here, enhancements are defined as the observed abundance of an element relative to C divided by the corresponding abundance relative to C in the solar corona as measured in gradual SEP events (Reames, 2014). Measured enhancements, and least-squares fits, are shown *vs. A/Q* at 2 and 4 MK for each event. Note that at 2 MK *A/Q* of O is similar to that of Ne, Mg and Si, while S has moved well above this group. At 4 MK the *A/Q* of Ca has moved well below that of Fe. Such changes affect the quality of the least-squares fit, $\chi^2$.

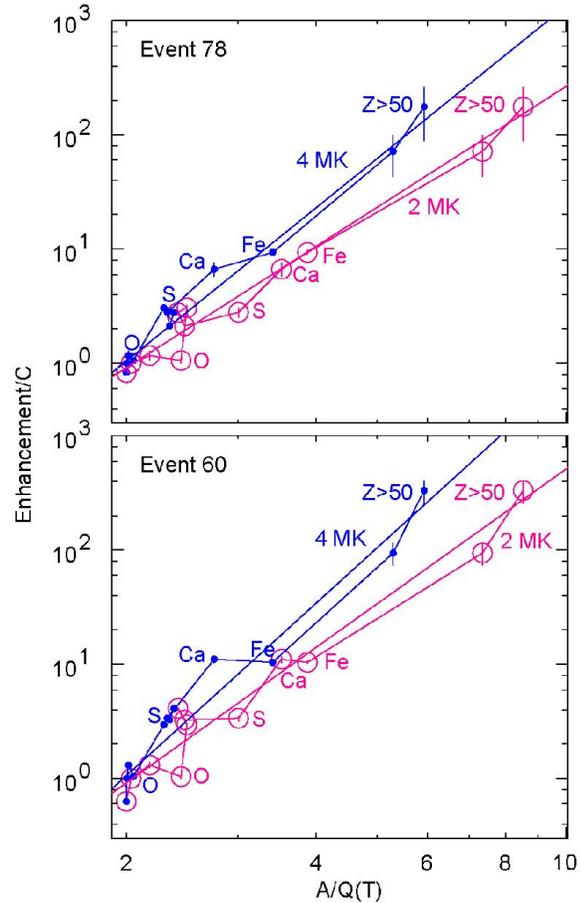

**Figure 2**. Enhancements/C (relative to coronal values as defined from gradual SEP events, *i.e.* Reames (2014)), are shown *vs. A/Q* for 2 MK (large red open circles) and 4MK (small blue circles) for two sample impulsive SEP events, events 60 (4 August 2002) and 78 (25 October 2003). Points, connected in order of Z, move horizontally with changing temperature by amounts that depend upon the species. He and C change little. As temperature increases, *A/Q* generally decreases as shown in Figure 1, but the spacing of *A/Q* varies with the element species.

## 2.2 Fitting Each Event

In fitting the events, we consider temperatures that are equally spaced in log *T*, namely 2, 2.5, 3.2, 4, and 5 MK. The procedure for determining the fit and temperature is as follows: For each event we choose a temperature, determine *A/Q* for each element (Figure 1) and obtain a linear least-squares fit to log (enhancement) *vs.* log (*A/Q*), noting the slope, *α*, and the goodness-of-fit, $\chi^2$. We then proceed to the next temperatures. The fit (and corresponding temperature) chosen for each event is that with the smallest $\chi^2$.

Values of $\chi^2$ vs. $T$ for all of the 111 SEP events from the list in Paper 1 are shown in Figure 3. The number of events having a minimum $\chi^2$ at each temperature is shown along the bottom of the figure. Values of 2.5 and 3.2 MK are clearly favored.

**Figure 3**. $\chi^2$ vs. $T$ is shown for all 111 impulsive SEP events using different colors and symbols for each event. The number of events with $\chi^2$ minima at each temperature is shown along the bottom of the panel.

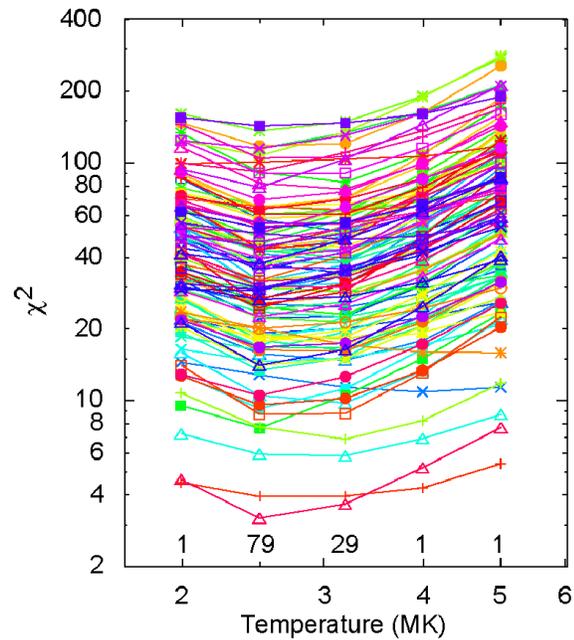

Thus, by choosing the minimum $\chi^2$ for each SEP event, we have determined both the best value of $\alpha$, and the most likely coronal temperature for that event. Some events have larger values of $\chi^2$ indicating errors in excess of those assumed. In weighting the measurements in the fits, for each species, we have assumed a fixed error of 15% in addition to the statistical error based on the number of ions sampled, since a scatter of 15–25% in abundances is seen even in well-measured events (see Section 2.3). It is possible that these fixed errors may represent event-to-event variations in the abundances of elements at different points in the corona sampled by different events at the time of acceleration.

We show the enhancement observations in the left panel of Figure 4 and the fit lines in the right panel. Observed enhancements with errors as large as the values are suppressed in the figure. Note that elements with $Z>26$ have a small effect on the fits owing to the small number of these ions in each event. Nevertheless, the fits seem to represent enhancements of these elements fairly well.

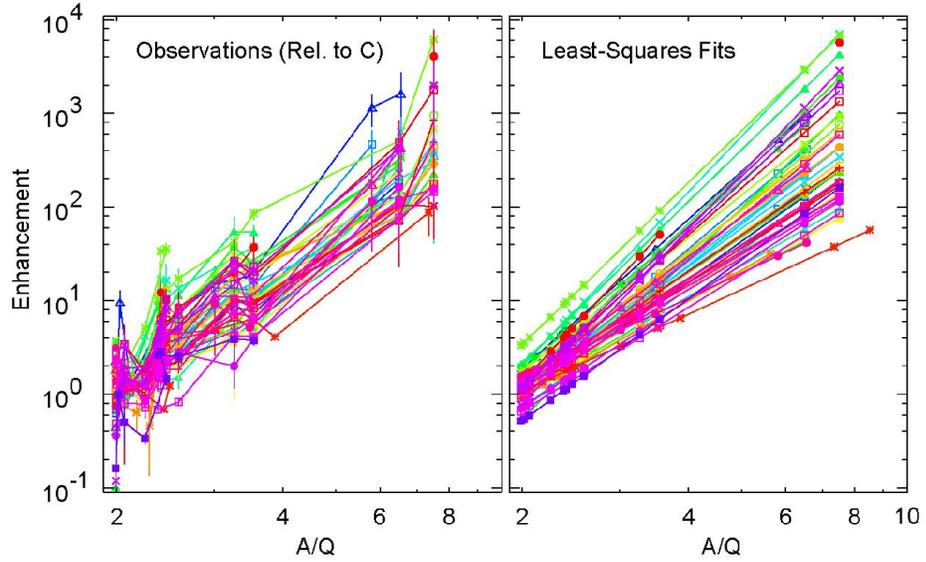

Figure 4. Enhancements/C, relative to the corona, are shown *vs. A/Q* in the left panel and as fitted lines in the right panel. Data from different SEP events are shown with differing colors and symbols.

The distribution of the fitted values of the power $\alpha$, where the enhancement $\sim (A/Q)^\alpha$ for SEP events with minimum $\chi^2$ at each temperature is shown in the histograms in Figure 5. Note that the events with $\alpha \geq 6$ come predominately from plasma at ~2.5 MK.

**Figure 5**. The panels show histograms of the distribution of the fitted values of $\alpha$, the power of the *A/Q* dependence of the abundance enhancement, for SEP events with a given value of the best-fit coronal temperature.

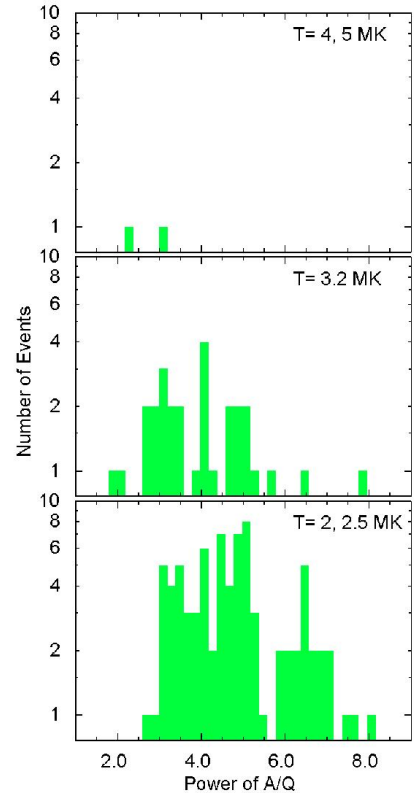

D. V. Reames, E. W. Cliver, S. W. Kahler

## 2.3 Residuals

Clearly the dominant behavior of the abundance enhancements is the power-law dependence upon *A/Q*. However, the residuals of these fits show some second-order behavior that is shown in Figure 6. Only measurements with a statistical error ≤ 20% are included so as to examine non-statistical variations. Points with values <1 fall below the fitted line, while those greater than one fall above, so that O and Fe fall below the fit by ~30%, Ne and Mg fall above by ~30%, while He, C, N, Si, S and fall closer to the fitted line. Thus the observed Ne/O is not completely explained by the fit.

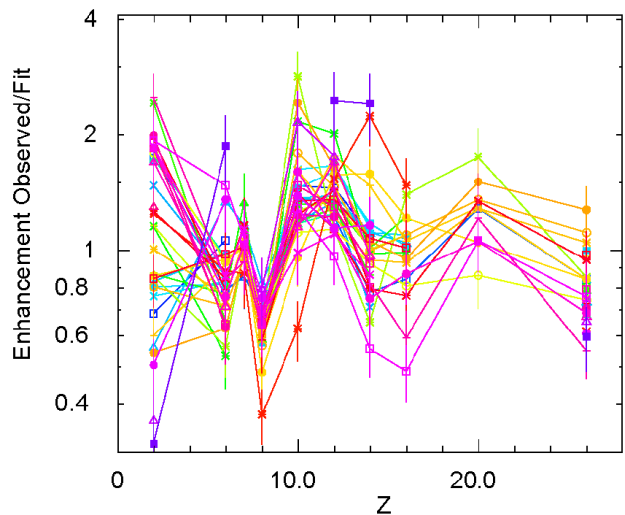

**Figure 6**. Residuals of the fits are shown as the observed enhancement divided by that at the fit line for each element. Only points with a statistical error of 20% or less are shown. For most elements the spread of the points is ~20% or more.

Part of this systematic underestimate of Ne/O and Mg/O by the fitting may come from the presence of other isotopes of Ne and Mg. We have assumed $^{20}$Ne and $^{24}$Mg in calculating *A/Q*, but recent measurements (Wiedenbeck *et al.*, 2010, see also Wiedenbeck *et al.*, 2008; Dwyer *et al.* 2001) in an impulsive SEP event show that $^{22}$Ne/$^{20}$Ne=0.321±0.018 and $^{26}$Mg/$^{24}$Mg=0.50±0.04. For the event observed by Wiedenbeck *et al.* (2010), our event 63 (20 Aug 2002), we find *α* = 4.92±0.71. Thus other Ne and Mg isotopes have *A/Q* values that are ~10% higher than we assumed, a significant difference, especially when raised to a large power, *α*. Heavier isotopes make a much smaller contribution for other elements.

In addition, Figure 6 shows that the spread of the points is generally larger than the ≤ 20% statistical error and depends somewhat on the element. This spread is especially large for He. In part this may come from background that is difficult to remove. In these Fe-rich events, the enhanced Fe stands out clearly from background from gradual events, for example, but often the unenhanced He (sometimes even depressed as "He-poor") does not. The unknown time dependence of any background makes it very difficult to remove. It is also possible that the abundance variations involve spatial variations in the underlying plasma probed by different SEP events.

## 3. Variation with CME Properties

We begin by comparing the new SEP parameters with the speed and width of the associated CMEs determined in Paper 1. This comparison is shown in Figure 7. In general, possible correlations are rather weak. However, the SEP events with steep enhancements ($\alpha \geq 6$) are preferentially associated with CME widths less than ~100 deg (11 of 11) and tend to have speeds less than ~800 km s$^{-1}$ (9 of 11).

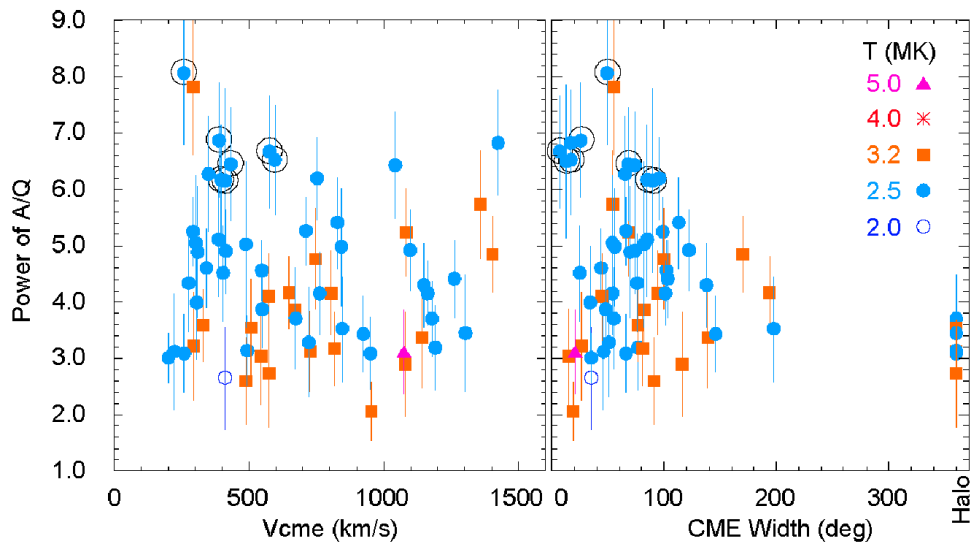

**Figure 7**. The power of the *A/Q* dependence of the SEP element enhancements, *α*, is shown as a function of the associated CME speed (left panel) and width (right panel). The derived coronal temperature is indicated by the shape and color of the plotted symbol as shown. Events identified as "He-poor" in Paper 1 are circled (see text).

In Paper 1 we identified as "He-poor" those events with He/O below about 40% of the coronal value. These events are circled in Figure 7. He-poor events are produced at low temperatures where *A/Q* of O is significantly above 2. In fact these events *all* have derived temperatures of 2.5 MK and all have $\alpha > 6$. The associated CMEs are narrow with speeds <700 km s$^{-1}$.

## 4. Variation with Flare X-ray Properties

While 90 of 111 SEP events have X-ray flare associations we limit our study to the 68 events for which solar longitudes (given in Table 2 of Paper 1) are determined and are measured to be <90º in order to exclude cases where some of the X-ray emission region is behind the solar limb. We consider the peak intensity of GOES 1–8 Å X-rays and its time integral, the X-ray fluence. The latter is shown for three sample events in Figure 8.

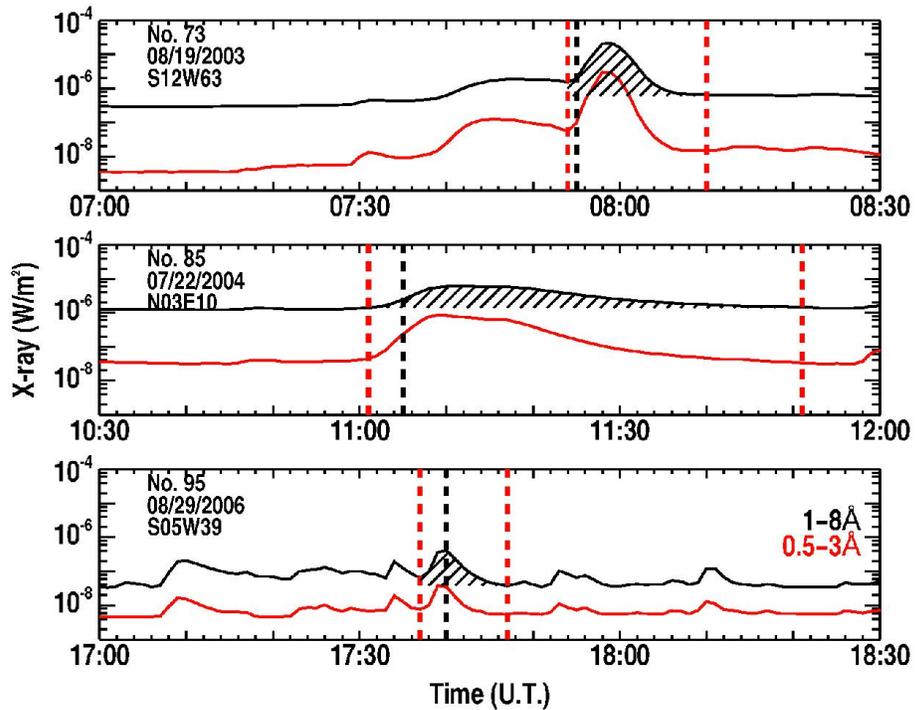

**Figure 8**. The intensities of 1-8 Å (black) and 0.5-3Å (red) X-rays are shown for three sample events. The shaded regions show the regions used to obtain the X-ray fluence for each event. The vertical black dashed line indicates the onset time of the type III radio burst and red dashed lines bound the X-ray emission.

Figure 9 shows $\alpha$, the power of the *A/Q* dependence of the abundance enhancements *vs.* the X-ray intensity, on the left, and the fluence on the right. The correlation coefficient for the intensity is -0.42 and for the fluence is -0.35. While these correlation coefficients are not large, the distribution of the SEP-determined temperatures is also interesting. Nearly all the flares of the B- and C-class or low fluence are 2.5 MK, while the flares of M- and X-class or high fluence have an unusually high proportion of 3.2 MK flares.

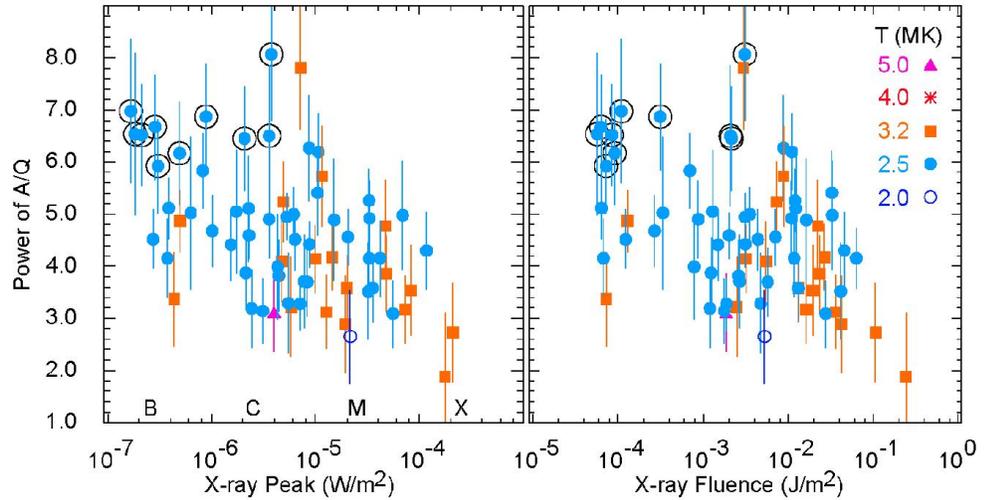

**Figure 9**. The power, α, of the element enhancement *vs. A/Q* in the impulsive SEP events is shown as a function of the peak X-ray intensity (left panel) or the X-ray fluence (right panel) of the associated flare. Symbol and color show the deduced SEP temperature. The X-ray class, B, C, M, or X, is shown along the bottom of the left panel. Events identified as "He-poor" in Paper 1 are circled (see text).

Again, in Figure 9 we circle events we identified as "He-poor" (Paper 1), those events with He/O below about 40% of the coronal value. He-poor events are produced at low temperatures where *A/Q* of O is significantly above 2. In fact these events *all* have derived temperatures of 2.5 MK and all have α ≥ 6. Here they are all B- and C-class flares.

We summarize the average values of α and temperature for SEP events with different X-ray flare classes in Table 1 and in Figure 9. Events with steep α tend to be associated with cooler (2.5 MK) plasma and smaller X-ray flares. Note that the X-ray flares associated with all of the He-poor events fall below class C4. The number of $^3$He-rich events, listed in Table 1, is discussed in Section 4.

Table 1. Variation of mean SEP properties with the X-ray flare class.

| X-ray class | B | C | M | X |
| --- | --- | --- | --- | --- |
| Events | 14 | 29 | 22 | 3 |
| Power, α | 5.28±0.26 | 4.38±0.17 | 4.16±0.19 | 3.36±0.57 |
| T (MK) | 2.60±0.06 | 2.71±0.09 | 2.76±0.08 | 2.97±0.19 |
| $^3$He/$^4$He≥0.1 | 11 | 15 | 7 | 0 |

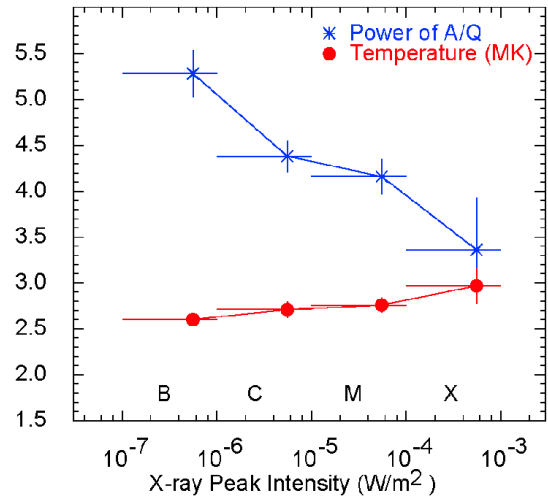

**Figure 10**. Variation in the mean value of *α*, the power of the element enhancement *vs. A/Q*, and of the coronal temperature is shown for SEP events with associated B-, C-, M-, and X-class X-ray flares.

It is not surprising that the X-ray intensity increases with the SEP-derived plasma temperature in these events. It is also possible to derive a temperature from fitting a thermal distribution to soft X-ray spectral data (e.g. Garcia, 1994). However, these latter temperatures are usually in the range 10–30 MK, temperatures where Ne would be *fully* ionized (see Figure 1), making it difficult to enhance Ne/O when both species have *A/Q*=2. Thus we conclude that the SEP-derived ionization-state temperatures of Table 1 and Figure 10 are more realistic measures of the actual electron temperatures in the coronal plasma where ion acceleration occurs. Probably the X-ray spectra are not entirely thermal.

As a final comment we should also mention the correlation of other SEP properties with the spectral indices. The correlation coefficient of *α* with the O spectral index is -0.20 (harder spectra have smaller *α*) while that of the log of the O fluence with spectral index is 0.42 (higher fluence means flatter spectra). Thus high-fluence SEP events tend to have harder spectra and somewhat more modest heavy-element enhancements.

## 5. $^3$He-rich Events

Resolution of isotopes of He in LEMT limits measurements of $^3$He/$^4$He abundances to ratios above ~0.1. Since this is 250 times the abundance in the solar wind (*e.g.* Gloeckler and Geiss, 1998), it is quite possible, even likely, that all of the impulsive events we study are somewhat $^3$He-rich. Here, however, we examine the subset of events with $^3$He/$^4$He ≥ 0.1 at 2–5 MeV/amu and, for the purposes of this paper, define this subset as $^3$He-rich. This $^3$He-rich subset is quite adequate to contrast their behavior with that of less $^3$He-rich SEP events.

As we discussed in the Introduction, enhancements in $^3$He/$^4$He are uncorrelated with those in Fe/C or Fe/O (*e.g.* Mason *et al.*, 1986; Reames, Meyer, and von Rosenvinge, 1994; however, see also Dwyer *et al.*, 2001), so $^3$He-rich events must be considered separately. The lower panel of Figure 11 shows $^3$He/$^4$He as an increasing circle size and varying color on a plot of the enhancement of Ne/C *vs.* Fe/C for our 111 events. The upper panel contrasts the behavior of *α*, the power of enhancement *vs. A/Q*, as the symbol size and color on an identical plot of Ne/C *vs.* Fe/C.

**Figure 11**. Each panel is a plot of the enhancements of Ne/C *vs.* Fe/C. In the lower panel the circle size and color varies with $^3$He/$^4$He as shown on the right-hand scale. In the upper panel the circle size and color varies with *α* as shown on the right-hand scale. As expected, *α* varies rather smoothly with enhancement, while $^3$He/$^4$He does not.

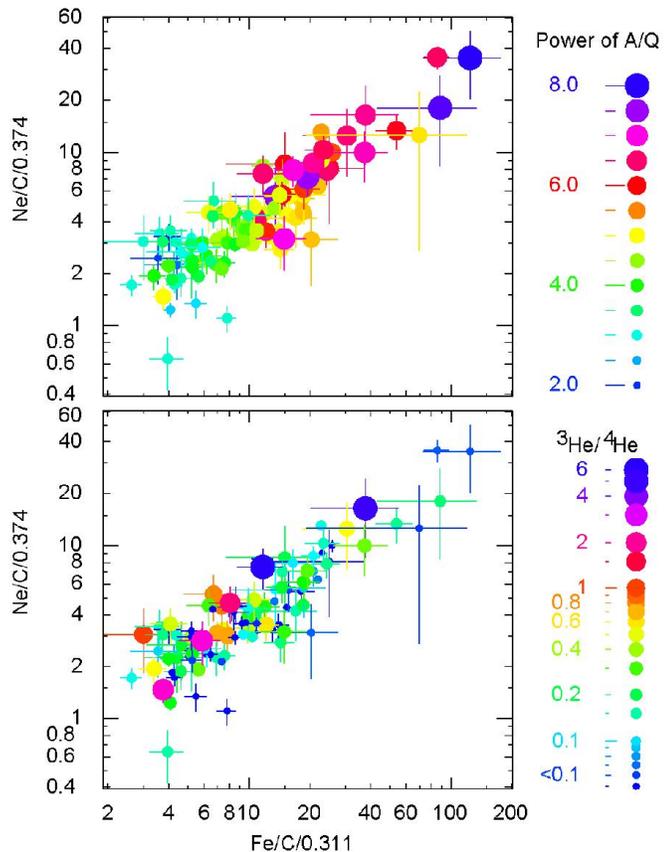

As we might expect, the parameter $\alpha$ increases smoothly with increasing enhancements since it was derived from these enhancements. In addition, the abundances of Ne/C and Fe/C are, themselves, highly correlated. However, $^3$He/$^4$He varies throughout. Events with the 100-fold enhancements of Fe/C and 40-fold Ne/C can be $^3$He poor while those with the lowest enhancements of Fe and Ne can be $^3$He rich or poor.

In Figure 12 we examine where the $^3$He-rich events fall within our measures of the associated CME and soft X-ray events. The $^3$He-rich events favor narrow CMEs with low speeds much more strongly than the $^3$He-poor events. They also favor small, B- and C-class X-ray flares, although they are seen from M-class flares. By contrast the $^3$He-poor events strongly favor the M-class flares. The X-ray distribution of $^3$He-rich events has been noted during previous solar cycles (Reames *et al.*, 1988).

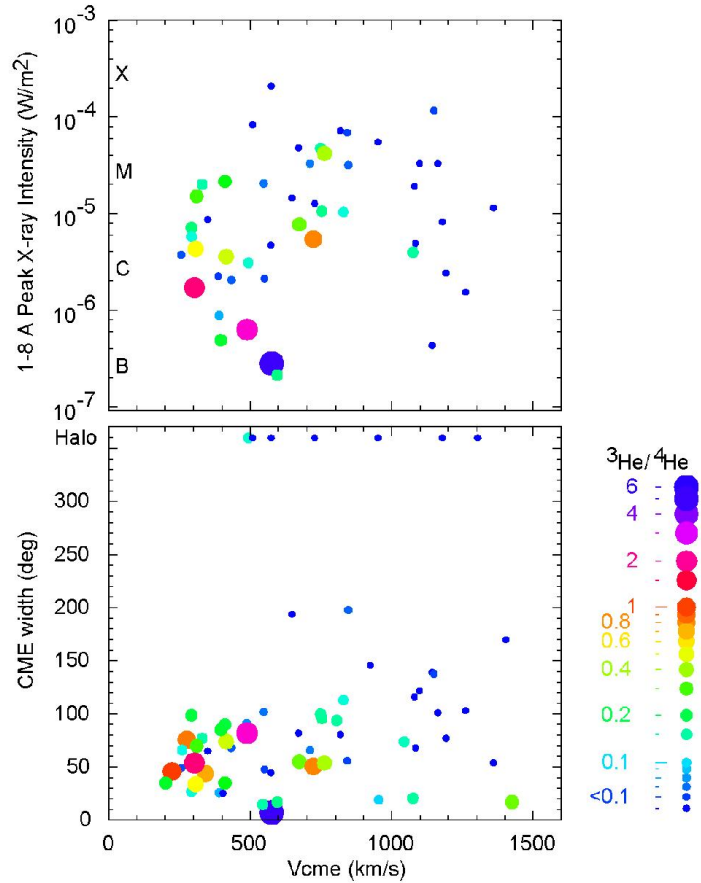

**Figure 12**. The figure shows the distribution of $^3$He/$^4$He, shown as varying symbol size and color, on plots of the CME width *vs.* CME speed (lower panel) and the soft X-ray intensity *vs.* CME speed (upper panel). $^3$He-rich events tend to dominate the lower left corner of both panels.

The number of $^3$He-rich events in each X-ray class is shown in Table 1. Of our total 111 Fe-rich SEP events, 68 have associated X-ray flares that are clearly on the disk and 33 of these have $^3$He/$^4$He $\geq$ 0.1. For B-, C-, and M-class flares, 11/14=79%, 15/29=51%, and 7/22=32%, respectively, are $^3$He-rich. Thus, $^3$He-rich SEP events are more strongly associated with smaller flares than are Fe-rich flares generally. In addition, only 11/61=18% of the events with $^3$He/$^4$He ≥ 0.1 have derived flare temperatures ≥ 3.2 MK while 50/61=82% have temperatures of 2.0–2.5 MK.

There are 32 $^3$He-rich events in our sample that have associated CMEs. Of these, 30/32=94% have widths ≤ 100° and 26/31=84% have speeds ≤800 km s$^{-1}$. Thus, $^3$He-rich events are even more strongly associated with narrow, slow CMEs and with B- and C-class X-ray flares than are other Fe-rich impulsive SEP events with lesser enhancements of $^3$He/$^4$He.

## 6. Discussion

*A priori*, the coronal temperature range leading to the observed abundances could have been bracketed easily by recourse to Figure 1. Above ~8 MK, Ne becomes fully ionized so both Ne and O have *A/Q*=2 and the observed enhancements in Ne/O would be difficult to produce. Below 2 MK, O acquires two electrons and joins Ne and Mg at *A/Q* ≈ 2.4, and again enhancement of Ne/O is problematic. However, events with lower SEP temperatures, where the pattern of *A/Q* becomes more complicated, do exist. Mason, Mazur, and Dwyer (2002) observed an event with enhanced N; they suggested it must have come from a region with a temperature below 1.5 MK. This small event was only observed at energies below 1 MeV amu$^{-1}$ and was not even seen by LEMT, suggesting a steep energy spectrum. As we have seen, small SEP events with steep energy spectra tend to have lower plasma temperatures. However, SEP events with temperatures below 2.0 MK do appear to be rare. *In general, the temperature scenario of Figure 1 combined with the power-law dependence on A/Q describes the observed pattern of abundances extremely well, from the extreme enhancements of Z>50, to the out-of-sequence enhancement of Ne/O to the He-poor events.*

Direct measurements of $Q_{Fe}$ of ions in the 0.18–0.43 MeV amu$^{-1}$ range during impulsive events show values in the range of 15–18 (DiFabio *et al.*, 2008) which might correspond to temperatures in the range 2–6 MK. The higher ionization states and evidence that elements up to Mg may be fully ionized appear to conflict with the enhancements of Ne/O. However, this higher ionization and especially the observed energy dependence of $Q_{Fe}$ (DiFabio *et al.* 2008) could easily result from additional stripping of the ions if they pass through very small amounts of matter after acceleration.

There is a weak tendency for events with steep energy spectra to have steep enhancements *vs. A/Q* (*i.e.* large *α*). This might suggest that abundance enhancements at SEP energies are large simply because high-Z elements have flatter energy spectra than those at lower *Z*. Some slight evidence of this is seen comparing Fe and O spectral indices (Paper 1). However, the mean value of *α* at 0.375 MeV amu$^{-1}$ is 3.26 (Mason *et al.*, 2004) while that at 3–10 MeV amu$^{-1}$ is 3.64±0.15 (Paper 1), hardly a large variation of enhancement with energy.

Note that enhancements of individual elements could vary from event to event even if *α* were constant, since *A/Q* varies with temperature. Decreasing temperatures generally increases *A/Q* which would increase enhancements. However, our analysis indicates that *both α and temperature* vary from event to event. As temperature decreases, *α* tends to increase, causing even larger enhancements than those produced by either variable alone.

It is rather surprising that nearly all of the events fall in such a narrow temperature window of 2.5–3.2 MK. This is driven mainly by the Ne/O enhancements and Ne/O was *not* considered in event selection. Only Fe/O was considered and Fe/O enhancements should occur over the entire coronal temperature region where Fe and O have significantly different values of *A/Q* (see Figure 1). The temperature range of 2.5–3.2 MK is just below the of the active region core temperature of ~4 MK (Warren, Winebarger, and Brooks 2012).

Finally, we should note that the correlations between SEPs, flares and CMEs are not terribly strong. However, this should *not* be surprising. That part of a magnetic reconnection on open field lines producing SEPs might be weakly correlated with that on nearby closed loops contributing to a flare. Any related mass ejection might also be variable as to spatial size and ejection speed. Perhaps it is surprising that any correlations remain. It is gratifying to find *any* new relationships, however limited, and to be able to quantify the event-to-event variations in abundance enhancements with plasma temperatures.

**Acknowledgments**: S.K. and E.C. were funded by AFOSR Task 2301RDZ4. We thank Alan G. Ling for preparing the X-ray fluences for our events.

Abundance Enhancement Variations in Impulsive SEP Events